\documentclass[preprint,showpacs,preprintnumbers,amsmath,amssymb,nofootinbib]{revtex4}
\usepackage{amsfonts}
\usepackage{booktabs}
\usepackage{mathrsfs}
\usepackage{epsfig}
\usepackage{graphicx}% Include figure files
\usepackage{dcolumn}% Align table columns on decimal point
\usepackage{bm}% bold math
\usepackage{amsmath}
\usepackage{slashed}

\let\jnfont=\rm
\def\NPB#1,{{\jnfont Nucl.\ Phys.\ B }{\bf #1},}
\def\PLB#1,{{\jnfont Phys.\ Lett.\ B }{\bf #1},}
\def\EPJC#1,{{\jnfont Eur.\ Phys.\ Jour.\ C }{\bf #1},}
\def\PRD#1,{{\jnfont Phys.\ Rev.\ D }{\bf #1},}
\def\PRL#1,{{\jnfont Phys.\ Rev.\ Lett.\ }{\bf #1},}
\def\MPLA#1,{{\jnfont Mod.\ Phys.\ Lett.\ A }{\bf #1},}
\def\JPG#1,{{\jnfont J.\ Phys.\ G}{\bf #1},}
\def\CTP#1,{{\jnfont Commun.\ Theor.\ Phys.\ }{\bf #1},}
\def\ZPC#1,{{\jnfont Z.\ Phys.\ C }{\bf #1},}
\def\JHEP#1,{{\jnfont JHEP \ }{\bf #1},}
\def\Rv{\not{\hbox{\kern-1pt $R$}}}
\def\p{\not{\hbox{\kern-3pt $p$}}}

\begin{document}
\preprint{\parbox{1.2in}{\noindent arXiv:1203.0694}}

\title{Higgs decay to dark matter in low energy SUSY:
       is it detectable at the LHC ? }

\author{Junjie Cao$^{\dag,\ddag}$, Zhaoxia Heng$^\dag$, Jin Min Yang$^\S$, Jingya Zhu$^\S$
        \\~ \vspace*{-0.3cm} }
\affiliation{
$^\dag$ Physics Department, Henan Normal University, Xinxiang 453007,  China\\
$^\ddag$ Center for High Energy Physics, Peking University,
       Beijing 100871, China \\
$^\S$ State Key Laboratory of Theoretical Physics, Institute of Theoretical Physics,
      Academia Sinica, Beijing 100190, China
     \vspace*{1.5cm}}

\begin{abstract}
Due to the limited statistics so far accumulated in the Higgs boson search at the LHC,
the Higgs boson property has not yet been tightly constrained and
it is still allowed for the Higgs boson to decay invisibly to dark matter
with a sizable branching ratio.
In this work, we perform a comparative study for the Higgs decay to neutralino 
dark matter by considering three different low energy SUSY
models: the minimal supersymmetric standard model (MSSM),
the next-to-minimal supersymmetric standard models (NMSSM) and the nearly minimal
supersymmetric standard model (nMSSM). Under current experimental constraints
at $2\sigma$ level (including the muon $g-2$ and the dark matter relic density), we scan
over the parameter space of each model. Then in the allowed parameter space we calculate
the branching ratio of the SM-like Higgs decay to neutralino dark matter and examine its
observability at the LHC by considering three production channels: the weak boson
fusion $VV\to h$, the associated production with a Z-boson $pp\to hZ+X$ or a pair of top
quarks $pp\to ht\bar{t}+X$.
We find that in the MSSM such a decay is far below the detectable level;
while in both the NMSSM and nMSSM the decay branching ratio can be large enough to
be observable at the LHC.
We conclude that at the LHC the interplay of detecting such an invisible decay
and the visible di-photon decay
may allow for a discrimination of different SUSY models.
\end{abstract}
\maketitle

\section{INTRODUCTION}
As a cornerstone of the standard model (SM) and also the last undiscovered piece,
the Higgs boson has been intensively searched in collider experiments.
The foregone colliders LEP II and Tevatron yielded null search
results, setting a lower bound of 114.4 GeV on the Higgs mass
\cite{h-hmB-LEP} and excluding a Higgs boson with a mass
around $2M_W$ \cite{tevatron-higgs}, respectively.
The ongoing Large Hadron Collider (LHC) took over the Higgs-hunting task
and recently reported its search results.
Based on an integrated luminosity of 4.9 $fb^{-1}$ collected at $\sqrt{s} = 7 TeV$,
the two experimental groups at the LHC independently further narrowed down the Higgs mass
region (at $95\%$ C.L. the CMS collaboration excluded 127-600 GeV while the ATLAS
collaboration excluded 112.9-115.5 GeV, 131-238 GeV and 251-466 GeV)
and both hinted to a Higgs boson around 125 GeV \cite{ATLAS-CMS-1112}.
Such a finding has stimulated some theoretical studies for
a Higgs boson near 125 GeV in low energy supersymmetry  \cite{h-125-susy}
and other models \cite{h-125-other}.

Of course, if the LHC hint of a 125 GeV Higgs from the di-photon channel is confirmed
in the future, it would severely constrain or exclude those new physics models in which
some new exotic decay modes (such as decaying invisibly into dark matter) are open and
the di-photon rate is suppressed. But so far the statistics at the LHC is
too small to confirm such a Higgs, let alone the precision measurement
of the Higgs decay branching ratios.
Therefore, experimentally it is still allowed for the Higgs boson to decay exotically,
such as invisibly to dark matter, with a sizable branching ratio
\footnote{In \cite{h-xx-gg-MSSM} the authors used the limited LHC statistics
of the di-photon signal rates to set constraints
on the invisible Higgs decay in the MSSM and found
that the invisible Higgs decay branching ratio around $10\%$ is allowed.}.

Theoretically, the Higgs decay to dark matter can indeed occur in some
new physics models, such as the gauge singlet extensions of the SM
\cite{h-xx-SSM},
the SM with a heavy fourth generation \cite{h-xx-4th},
the large extra dimension model \cite{h-xx-EDM}, the technicolor model \cite{h-xx-TC},
the spontaneously broken R-parity models \cite{h-xx-RPV} and the non-linearly
realized supersymmetric model \cite{h-xx-NLS} and the MSSM with a singlet \cite{h-xx-SMSSM}.
In this work, we perform a comparative study for the Higgs decay to neutralino dark matter 
in low energy SUSY
by considering three different models: the minimal supersymmetric standard model (MSSM)
\cite{MSSM-old, MSSM-Martin, h-MSSM-Djouadi},
the next-to-minimal supersymmetric standard models (NMSSM) \cite{NMSSM-rev, NMSSM-ph}
and the nearly minimal supersymmetric standard model (nMSSM) \cite{nMSSM-Wagner,MNSSM-Neu,09-nMSSM-Cao}.
As will be shown, in both the NMSSM and nMSSM, the SM-like Higgs boson can decay to
neutralino dark matter with a sizable branching ratio.

In case that the Higgs boson decays to dark matter with a sizable branching ratio,
detecting such a decay at the LHC will be important because in this case the conventional
visible decays into $\gamma \gamma$, $b\bar b$, $\tau \bar\tau$, $WW^{(*)}$ and
$ZZ^{(*)}$ are often suppressed.
Obviously, the main production channel via gluon fusion $gg \to h$ is not usable
because it just gives missing energy.
It was found through Monte Carlo simulations that the production
via vector boson fusion (VBF) $pp \to hqq'$ and the associated productions
$pp \to hZ$ and $pp \to ht \bar{t}$ can offer the opportunity to
detect the Higgs decay to dark matter
\cite{h-xx-MC-VBF,h-xx-MC-hZ,h-xx-HanTao,h-xx-MC-hZ-hW,h-xx-MC-htt}.
So in this work we choose these three production channels to display 
the observability of Higgs decay to dark matter in  low energy SUSY.

Note that although in the literature the Higgs decay to nutralino dark matter has been 
discussed in some specific model like MSSM, it is necessary to give a
revisit in low energy SUSY for two reasons: 
(i) Different SUSY models usually give rather different
phenomenology and it is interesting to perform a comparative study for different models;
(ii) We want to know in the SUSY parameter space allowed by current experiments     
whether or not the Higgs decay to nutralino dark matter is detectable at the LHC. 
If in some model this decay is found to be accessible at the LHC, 
we further want to know how large the parameter space can be covered by searching
for such an invisible decay at the LHC.  

This work is organized as follows. In Sec. II, we briefly describe the three
supersymmetric models. In Sec. III, through a scan over the parameter space we
present the branching ratio of Higgs decay to
neutralino dark matter and show the observability at the LHC.
Finally, the conclusion is given in Sec. IV.

\section{The SUSY models}
In a renormalizable supersymmetric field theory, the interactions and masses of all particles
are determined by their gauge transformation properties and superpotential.
The superpotential is a holomorphic function of chiral superfields
$\hat{\Phi}_{i}\supset(\phi_{i}, \psi_{i}, F_{i})$,
with $\phi_{i}$, $\psi_{i}$ and $F_{i}$ being respectively the bosonic, fermionic
and auxiliary fields, and takes a form \cite{MSSM-Martin}
\begin{eqnarray}
W &=& L^{i}\hat{\Phi}_{i}+\frac{1}{2}M^{ij}\hat{\Phi}_{i}\hat{\Phi}_{j}
      +\frac{1}{6}y^{ijk}\hat{\Phi}_{i}\hat{\Phi}_{j}\hat{\Phi}_{k}
\end{eqnarray}
where the parameters $L_{i}$ should be of dimension $[mass]^{2}$ and is only allowed
if $\hat{\Phi}_{i}$ is a gauge singlet. The mass matrix $M^{ij}$ can only be non-zero
when the supermultiplets $\hat{\Phi}_{i}$ and $\hat{\Phi}_{j}$ are conjugates of
each other under gauge transformation. And the massless coefficients $y^{ijk}$ can
only be non-zero when $\hat{\Phi}_{i}\hat{\Phi}_{j}\hat{\Phi}_{k}$ formed a gauge singlet.

The MSSM is the most economized realization of supersymmetry in particle physics, which
has two Higgs doublets $\hat{H}_u$ and $\hat{H}_d$ and its superpotential is given by \cite{MSSM-Martin}
\begin{eqnarray}
W^{MSSM}=W_{F}+\mu\hat{H}_{u}\cdot\hat{H}_{d},
\end{eqnarray}
with $W_{F}$ given by
\begin{eqnarray}
W_{F}=\overline{u}Y_{u}\hat{Q}\cdot\widehat{H}_{u}-\overline{d}Y_{d}\hat{Q}\cdot\hat{H}_{d}
-\overline{e}Y_{e}\hat{L}\cdot\hat{H}_{d} .
\end{eqnarray}
The MSSM has the so-called $\mu$-problem, which can be solved in some extensions
by introducing a Higgs singlet $\hat{S}$. Among these extensions the most popular
ones are the NMSSM and nMSSM, whose superpotentials are \cite{NMSSM-rev, nMSSM-Wagner}
\begin{eqnarray}
W^{NMSSM}&=&W_{F}+\lambda\hat{S}\hat{H}_{u}\cdot\hat{H}_{d}+\frac{\kappa}{3}\hat{S}^{3};\\
W^{nMSSM}&=&W_{F}+\lambda\hat{S}\hat{H}_{u}\cdot\hat{H}_{d}+\xi_{F}M^{2}_{\rm n}\hat{S}.
\end{eqnarray}
The scalar potential in the Lagrangian contains the so-called F-term, D-term and soft-term \cite{Barger-06-Emssm}:
\begin{eqnarray}%\mathscr{L}
V^{SUSY}=V_{F}+V_{D}+V_{soft},
\end{eqnarray}
where
\begin{eqnarray}
&&V_{F}=F^{*i}F_{i}, F^{*i}=-W^{i}=-\frac{\delta W}{\delta \hat{\Phi}_{i}};\\
&&V_{D}=\frac{G^{2}}{8}(|H_d|^{2}-|H_u|^{2})^{2}+\frac{g^{2}_{\rm 2}}{2}(|H_{d}|^{2}|H_{u}|^{2}
-|H_{u}\cdot H_{d}|^{2});\\
&&V^{MSSM}_{\rm soft}=\tilde{m}^{2}_{\rm H_{u}}|H_{u}|^{2}+\tilde{m}^{2}_{\rm H_{d}}|H_{d}|^{2}
+(B_{\mu}H_{u}\cdot H_{d}+h.c.);\\
&&V^{NMSSM}_{\rm soft}=\tilde{m}^{2}_{\rm H_{u}}|H_{u}|^{2}+\tilde{m}^{2}_{\rm H_{d}}|H_{d}|^{2}
+\tilde{m}^{2}_{\rm S}|S|^{2}+(\lambda A_{\lambda}SH_{u}\cdot H_{d}
+\frac{\kappa}{3}A_{\kappa} S^{3}+h.c.)\\
&&V^{nMSSM}_{\rm soft}=\tilde{m}^{2}_{\rm H_{u}}|H_{u}|^{2}+\tilde{m}^{2}_{\rm H_{d}}|H_{d}|^{2}
+\tilde{m}^{2}_{\rm S}|S|^{2}+(\lambda A_{\lambda}SH_{u}\cdot H_{d}+\xi_{S}M^{2}_{\rm n}\hat{S}+h.c.)
\end{eqnarray}
Here the parameter $G$ is defined as $G^2=g_1^2+g_2^2$ with $g_1$ and $g_2$ 
denoting respectively the $U(1)_Y$ and $SU(2)_L$ couplings. 
With the Higgs fields $H_{u}$, $H_{d}$ and $S$ developing respectively the VEV $v_{u}$ $v_{d}$
and $v_{S}$, they can be rewritten as
\begin{equation}
H_{u}=\left(        \begin{array}{c}
          H^{+}_{\rm u} \\
          \frac{v_{u}+\phi_{u}+i\varphi_{u}}{\sqrt{2}} \\
        \end{array} \right) ,
H_{d}=\left(         \begin{array}{c}
          \frac{v_{d}+\phi_{d}+i\varphi_{d}}{\sqrt{2}} \\
          H^{-}_{\rm d} \\
         \end{array}        \right)   ,
S=\frac{v_{S}+\phi_{S}+i\varphi_{S}}{\sqrt{2}}
\end{equation}
In both the NMSSM and nMSSM we have five complex scalar fields or ten real scalar degrees
of freedom, whose mass eigenstates are obtained as
\begin{eqnarray}
\left( \begin{array}{c}
    h_{1} \\ h_{2} \\ h_{3} \\
  \end{array} \right)
= S_{ij} \left( \begin{array}{c}
    \phi_{u} \\ \phi_{d} \\ \phi_{S} \\
  \end{array} \right),
\left( \begin{array}{c}
    a_{1} \\ a_{2} \\ G^{0} \\
  \end{array} \right)
=P_{i,j} \left(  \begin{array}{c}
    \varphi_{u} \\ \varphi_{d} \\ \varphi_{S} \\
  \end{array} \right) ,
\left( \begin{array}{c}
    H^{+} \\ G^{+} \\
  \end{array} \right)
=C_{ij} \left(  \begin{array}{c}
    H^{+}_{\rm u} \\ H^{+}_{\rm d} \\
  \end{array} \right)
\end{eqnarray}
Here the three Goldstone bosons $G^0$ and $G^\pm$ will be eaten by the weak gauge bosons
$Z$ and $W^{\pm}$ respectively. Then we have seven Higgs bosons, among which
$h_{1}$, $h_{2}$ and $h_{3}$ are CP-even (with the convention $m_{h_{1}}<m_{h_{2}}<m_{h_{3}}$),
$a_{1}$ and $a_{2}$ are CP-odd (ordered as $m_{a_{1}}<m_{a_{2}}$), and $H^{\pm}$ are the charged
ones.

In the NMSSM and nMSSM there are five neutralinos ($\chi^0_{\rm i}$),
which are the mixture of bino ($\tilde{B}$), wino ($\tilde{W^{0}}$),
higgsino ($\tilde{H_{u}}$, $\tilde{H_{d}}$) and singlino ($\tilde{S}$):
\begin{eqnarray}
\left(  \begin{array}{c}
    \chi^0_{\rm 1} \\ \chi^0_{\rm 2} \\
    \chi^0_{\rm 3} \\ \chi^0_{\rm 4} \\
    \chi^0_{\rm 5} \\
  \end{array} \right)
=N_{ij} \left(  \begin{array}{c}
    \tilde{B} \\     \tilde{W^{0}} \\
    \tilde{H_{u}} \\    \tilde{H_{d}} \\
    \tilde{S} \\ \end{array} \right)
\end{eqnarray}
We assume the lightest neutralino is the lightest supersymmetric particle (LSP) and make
up of the cosmic dark matter.

For the purpose of our numerical analysis, we present the interactions of the Higgs
bosons $h_i$ with $b\bar{b}$, $\tau\tau$, $WW$ and $\chi^0_{\rm 1}\chi^0_{\rm 1}$. In the
the NMSSM, they are given by \cite{h-NMHDecay-Ulrich}
\begin{eqnarray}
&&h_{i}b_{L}b^{c}_{\rm R}:\ \ \ \ ~\frac{m_{b}}{\sqrt{2}v\cos\beta}S_{i2} \\
&&h_{i}\tau_{L}\tau^{c}_{\rm R}: \ \ \ \ ~\frac{m_{\tau}}{\sqrt{2}v\sin\beta}S_{i1} \\
&&h_{i}W^{+}_{\rm \mu}W^{-}_{\rm \nu}:
\ g_{\mu\nu}\frac{g^{2}_{\rm 2}}{\sqrt{2}}(h_{u}S_{i1}+h_{d}S_{i2}) \\
&&h_{i}\chi^0_{\rm 1}\chi^0_{\rm 1}: \ \ \ \ \ \frac{\lambda}{\sqrt{2}}(S_{i1}\Pi^{45}_{\rm 11}
+S_{i2}\Pi^{35}_{\rm 11}+S_{i3}\Pi^{34}_{\rm 11})-\sqrt{2}\kappa S_{i3}N_{15}N_{15}\nonumber\\
&& \ \ \ \ \ \ \ \ \ \ \ \ \ \ \ -\frac{g_{1}}{2}(S_{i1}\Pi^{13}_{\rm 11}-S_{i2}\Pi^{14}_{\rm 11})
+\frac{g_{2}}{2}(S_{i1}\Pi^{23}_{\rm 11}-S_{i2}\Pi^{24}_{\rm 11}) \label{ghxx}
\end{eqnarray}
where $\Pi^{ij}_{11}=N_{1i}N_{1j}+N_{1j}N_{1i}$. The corresponding couplings in the nMSSM
can be obtained by setting $\kappa$ equal to zero (for MSSM, setting $\kappa$, $\lambda$,
$S_{13}$ and $N_{15}$ to zero).

For the NMSSM, in the basis
$\chi^{0}=(\tilde{B}, \tilde{W^{0}}, \tilde{H_{u}}, \tilde{H_{d}}, \tilde{S})$,
the tree-level neutralino mass matrix takes the form \cite{h-NMHDecay-Ulrich, NMSSM-rev}
\begin{eqnarray}
M_{\tilde{\chi}^{0}}= \left( \begin{array}{ccccc}
    M_{1} & 0     & \frac{g_{1}v_{u}}{\sqrt{2}} & -\frac{g_{1}v_{d}}{\sqrt{2}}    & 0 \\
    0     & M_{2} & -\frac{g_{2}v_{u}}{\sqrt{2}} & \frac{g_{2}v_{d}}{\sqrt{2}}    & 0 \\
 \frac{g_{1}v_{u}}{\sqrt{2}} & -\frac{g_{2}v_{u}}{\sqrt{2}}     & 0 & -\mu & -\lambda v_{d} \\
 -\frac{g_{1}v_{d}}{\sqrt{2}} & \frac{g_{1}v_{d}}{\sqrt{2}}   & -\mu & 0 & -\lambda v_{u} \\
    0     & 0                 & -\lambda v_{d}             & -\lambda v_{u} & 2\kappa S \\
  \end{array} \right)\label{Neu-Matric}
\end{eqnarray}
The corresponding mass matrix for the MSSM can be obtained by taking
the upper 4 $\times$ 4 matrix from the above expression, and for the nMSSM
it can be obtained by set the term $2 \kappa S$ to zero.
When $|\mu_{eff}|$ or $M_2 \gg M_{Z}$, the lightest neutralino in the MSSM
becomes bino-like, with a mass given by \cite{MSSM-Martin}
\begin{eqnarray}
m_{\chi ^{0}_{\rm 1}}\simeq M_{1}-\frac{m^{2}_{\rm z} \sin ^{2} \theta _{w} (M_{1}
+\mu \sin 2\beta)}{\mu ^{2}-M^{2}_{\rm 1}} \label{Mx-MSSM}
\end{eqnarray}
In the nMSSM the lightest neutralino is singlino-like and its mass can be
approximated as \cite{MNSSM-Neu}
\begin{eqnarray}
m_{\chi^{0}_{\rm 1}} \simeq \frac{2\mu \lambda^{2}(v^{2}_{\rm u}
+v^{2}_{\rm d})}{2\mu^{2}+\lambda^{2}(v^{2}_{\rm u}+v^{2}_{\rm d})}
\frac{\tan \beta}{\tan^{2}\beta+1}. \label{Mx-nMSSM}
\end{eqnarray}

\section{Numerical Results and Discussions}
%\subsection{Description of numerical calculation}
We scan over the parameter space of each model under current experiment constraints, and
for each survived sample we calculate the Higgs spectrum, decay branching ratios and production
rates at the LHC. In our calculation we use the package NMSSMTools 
\cite{h-NMHDecay-Ulrich} and
extend it to the nMSSM \cite{09-nMSSM-Cao}. 
For the calculation of $h\to \gamma\gamma$ in the SM,
we use the package Hdecay \cite{HDecay-Spira}.
For the Higgs production cross sections, we use the code on the website \cite{Spira-code}
(this code is aimed at the MSSM, and we extend it to the NMSSM and the nMSSM).
For parton distributions we use CTEQ6L \cite{PDF-CTEQ6L} with the renormalization scale
and the factorization scale chosen to be the sum of the masses of the produced particles.

In our scan we require the models to explain
the cosmic dark matter relic density measured by WMAP \cite{ex-DM}
and also explain the muon anomalous magnetic moment at $2\sigma$ level.
In addition, we consider the following experimental constraints:
\begin{enumerate}
    \item The LEP bounds on sparticle masses and on the Higgs sector from 
     $e^+e^-\rightarrow hZ (hA)$
          followed by $Z\rightarrow \ell^+ \ell^- ,\chi^0_1\chi^0_1$,
          $h\rightarrow b\bar{b}$ and $\tau^+\tau^-$ \cite{ex-LEP};
We also consider the LEP-I constraints on the invisible $Z$ 
decay \footnote{Note that the constraints from such an invisible $Z$ 
     decay are stringent for a wino-like or higgsino-like 
     neutralino, but become quite weak for a bino-like or singlino-like 
     neutralino.}, i.e., $\Gamma(Z\to \chi_1^0 \chi_1^0) < 1.76~{\rm MeV}$,
     and the LEP-II constraints on neutralino production
     $\sigma(e^+e^-\to \chi_1^0 \chi_i^0) < 10^{-2}~{\rm pb}~ (i>1)$
     and 
    $\sigma(e^+e^-\to \chi_i^0 \chi_j^0) < 10^{-1}~{\rm  pb}~ (i,j>1)$
     \cite{ex-LEP}.

    \item The Tevatron bounds on sparticle masses and on stop or sbottom pair production followed by
          $\tilde{t}\rightarrow c\chi,bl\tilde{v_{l}}$ and $\tilde{b}\rightarrow b\chi$
          \cite{ex-Tev, ex-Tev-2};
    \item The recent LHC bounds on the Higgs sector from the measurement of the signal rates of
          $\gamma\gamma$, $\tau\tau$, $WW^{(*)}$ and $ZZ^{(*)}$ \cite{ex-LHC-1, ex-LHC-2};
    \item The constraints from B-physics, such as $b\rightarrow s\gamma$ and
          $B_{s}\rightarrow \mu\mu$ \cite{ex-B};
    \item The electroweak precision observables like $M_{W}$, $\sin^{2}\theta_{W}$ and $R_b$
          \cite{ex-EW-pre}.
\end{enumerate}
In order to reduce the number of free parameters,
for the gaugino masses we assume the grand unification relation
$3M_{1}/5\alpha_{1} = M_{2}/ \alpha_{2} = M_{3}/ \alpha_{3}$ and thus we have
only one gaugino mass parameter (we choose $M_2$ in our calculation).
In order to explain the muon anomalous magnetic moment at $2\sigma$ level
for moderate $\tan \beta$ ($\leq$ 20),
for the smuon sector we assume the soft-breaking parameters to be 100 GeV \cite{09-nMSSM-Cao}.
For other soft-breaking parameters in the squark and slepton sectors,
we assume them to be 1 TeV except that for the MSSM we allow the third-generation
squark mass parameters $m_{\tilde{q}}$ to vary in a wide range.
The parameter $M_{n}$ in the superpotential of the nMSSM is also fixed to be 1 TeV.
Other parameters are scanned in the following ranges
(among which $M_{a}$ is the mass of the $\cos \beta \varphi_u + \sin \beta \varphi_d$ field,
i.e., the diagonal element of the doublet in the CP-odd Higgs mass matrix):
\begin{enumerate}
\item For the MSSM:
$1 < \tan \beta < 20$, $100 {\rm ~GeV}< \mu < 600 {\rm ~GeV}$,
$10 {\rm ~GeV} < M_2 < 200 {\rm ~GeV}$,
$500  {\rm ~GeV}< M_{a} < 3{\rm ~TeV}$,
$100 {\rm ~GeV}< M_{\tilde{q}} < 2 {\rm ~TeV}$,
$-3 {\rm ~TeV}< A_{t,b} <3 {\rm ~TeV}$.
In this space we scan two million random points and about
three thousands points survived the experimental constraints.
\item For the NMSSM:
$0.1 < \lambda < 0.7$,
$0.1 < \kappa < 0.5$,
$1 < \tan \beta < 4$,
$100 {\rm ~GeV}< (\mu,M_a)<  1 {\rm ~TeV}$,
$50 {\rm ~GeV}< M_2<  150 {\rm ~GeV}$,
$0 < A_{\lambda} < 1 {\rm ~TeV}$,
$-500 {\rm ~GeV} < A_{\kappa} < 0$.
In this space we scan ten million random points and about one thousand points survived.
\item For the nMSSM:
$0.1 < \lambda < 0.7$,
$1 < \tan \beta < 10$,
$50 {\rm ~GeV}< (\mu,M_2,M_a)<  1 {\rm ~TeV}$,
$0 < A_{\lambda} < 1 {\rm ~TeV}$,
$0 < M_{\tilde{S}} < 500{\rm ~GeV}$,
$-1 < \xi_{F} < 1$.
In this space we scan one billion random points and about 2 thousands points survived.
\end{enumerate}
Note that our scan is not a general scan over the entire parameter space.
Since our purpose is to figure out if the invisible decay of the Higgs boson
is accessible at the LHC, we only scanned over a part of parameter space which
is potentially able to allow for a light neutralino and hence the Higgs can
decay into the neutralino pair.

%\subsection{Numerical results and discussions}
%%%fig.1 %%%%%%%%%%%%%%%%%%%%%%%%%%%%
\begin{figure}[thb]
\epsfig{file=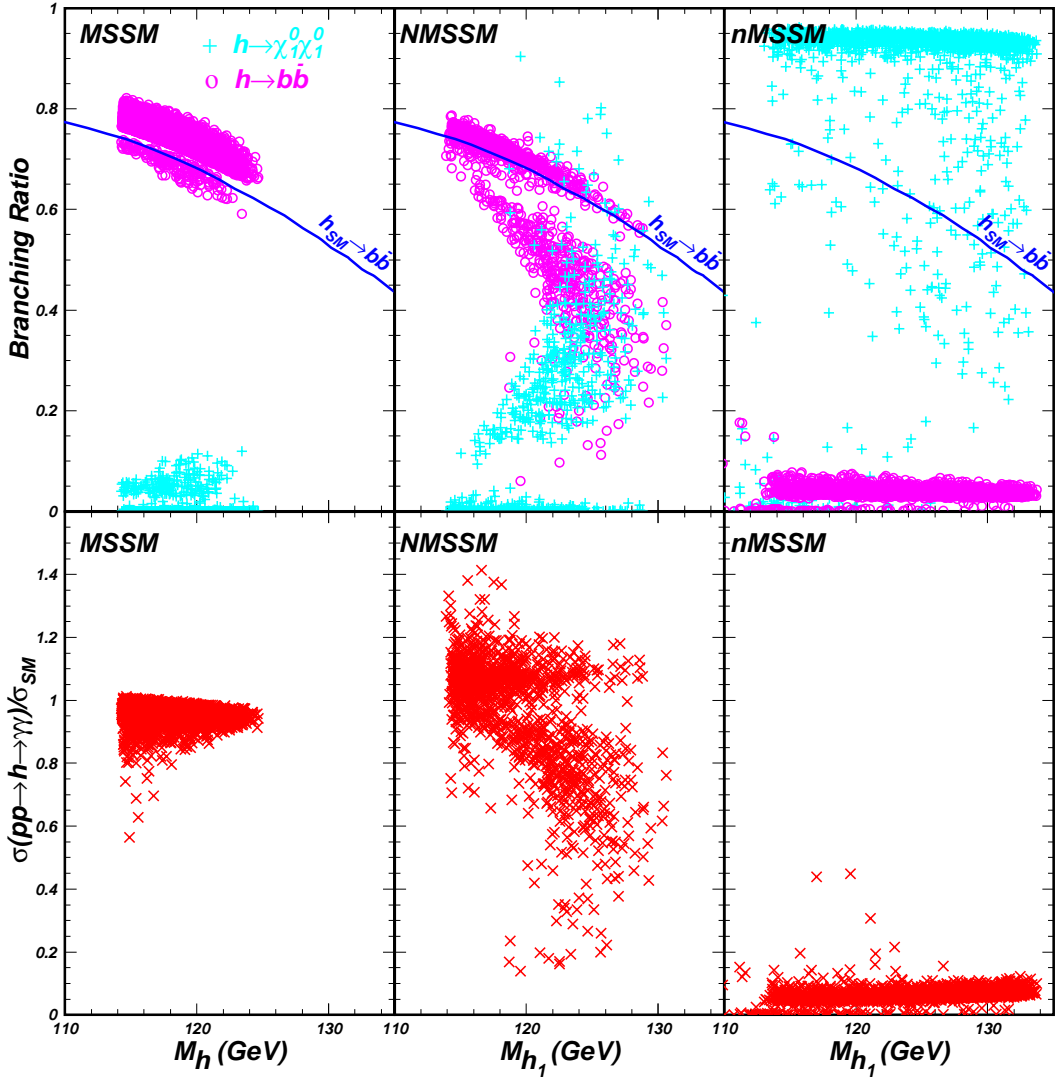,width=14cm} \vspace{0.3cm}
\vspace{-1cm}
\caption{The scatter plots of the parameter space which survive all constraints listed in the text.
In the upper frames, the samples denoted by crosses (sky-blue) show the branching ratio
of $h\to \chi_{1}^{\rm 0} \chi_{1}^{\rm 0}$ while
the samples denoted by circle (magenta) show the branching ratio of $h\to b \bar{b}$.
The solid curves (blue) denote the SM prediction for the branching ratio of  $h\to b \bar{b}$.
In the lower frames, the samples denoted by times (red) show the ratio
$\sigma_{SUSY}(pp \to h \to \gamma\gamma)/\sigma_{SM}(pp \to h \to \gamma\gamma)$
at the LHC (7 TeV). 
In this figure and below, '$h_{SM}$' denotes the Higgs boson in the SM,
'h' denotes the lightest nutral Higgs boson in the MSSM, 
and '$h_{1}$' denotes the SM-like Higgs boson in the NMSSM and nMSSM
(the doublet component of $h_1$ is over 60\%).}
\label{fig1}
\end{figure}
%%%%%%%%%%%%%%%%%%%%%%%%%%%%%%%%%%%%%
%%%fig.2 %%%%%%%%%%%%%%%%%%%%%%%%%%%%
\begin{figure}[thb]
\epsfig{file=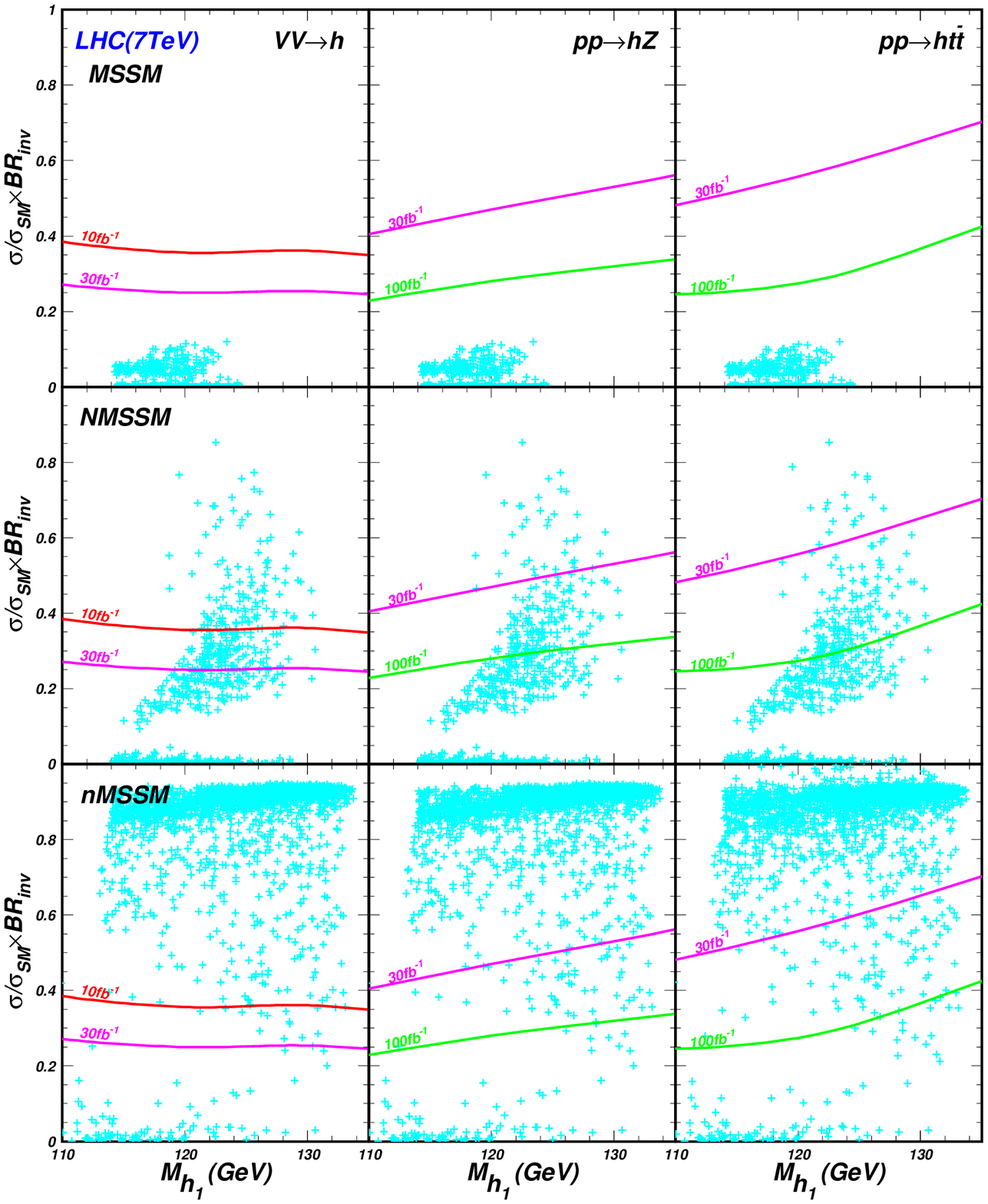,width=14cm} \vspace{0.3cm}
\vspace{-1cm}
\caption{Same as Fig.1, but showing the quantity
$\frac{\sigma_{SUSY}}{\sigma_{SM}}\times Br(h_1\to \chi_{1}^0 \chi_{1}^0)$ with $\sigma_{SUSY}$
($\sigma_{SM}$) being the SUSY (SM) Higgs production rates for the processes
$VV \to h$, $pp\to hZ$ and $pp\to ht \bar{t}$.
The solid curves show the $2\sigma$ sensitivity  \cite{h-xx-MC-VBF,h-xx-MC-hZ,h-xx-MC-htt}
of the ATLAS detector at the LHC (7 TeV) with 10 $fb^{-1}$, 30 $fb^{-1}$ and 100 $fb^{-1}$
(the region above each curve is the observable region).}
\label{fig2}
\end{figure}
%%%%%%%%%%%%%%%%%%%%%%%%%%%%%%%%%%%%%
In Figs.~\ref{fig1} and \ref{fig2} we display the scatter plots of the parameter space
which survive all constraints.
Fig.~\ref{fig1} shows the Higgs decay branching ratios and the LHC (7 TeV) di-photon rate,
while Fig.~\ref{fig2} shows the observability of the decay $h\to \chi_{1}^{\rm 0} \chi_{1}^{\rm 0}$
through three production channels at the LHC (7 TeV).
The $2\sigma$ sensitivity of the ATLAS detector shown in Fig.~\ref{fig2} is obtained
from a Monte Carlo simulation carried out in \cite{h-xx-MC-VBF,h-xx-MC-hZ,h-xx-MC-htt}.
For the decay $h\to \chi_{1}^{\rm 0} \chi_{1}^{\rm 0}$,
the signature of the production via vector boson fusion $VV\to h$
is two far forward and backward tagging jets of moderate $p_{T}$
with considerable missing momentum $\p_{T}$ in the central region.
For the productions $pp\to hZ$ and $pp\to h t\bar{t}$,
the signatures are obvious: for the former it is two isolated high $p_{T}$ leptons
from the Z-boson decay and large missing $\p_{T}$ from the
Higgs decay; for the latter it is di-leptons (or lepton plus jets) and
large missing $\p_{T}$.

From these figures we obtain the following findings:
\begin{itemize}
\item In each model the SM-like Higgs can have a mass near 125 GeV,
as hinted by the recent LHC results.
\item In the MSSM the SM-like Higgs boson dominantly decays to $b\bar{b}$ (just like in the SM),
the decay $h\to \chi_{1}^0 \chi_{1}^0$ has a very small branching ratio (below about $10\%$),
and the di-photon signal rate is close to the SM value.
Due to the small branching ratio, the Higgs decay  $h\to \chi_{1}^0 \chi_{1}^0$
is far below the detectable level.

\item In the NMSSM the decay $h_1\to \chi_{1}^0 \chi_{1}^0$ can be comparable to $h_1\to b\bar{b}$.
In the region with a sizable decay ratio of  $h_1\to \chi_{1}^0 \chi_{1}^0$,
the lightest neutralino $\chi_{1} ^{\rm 0}$ is rather light (below $h_{1}/2$) and the coupling
of $h_{1}$ to $\chi_{1} ^{\rm 0}$ is large (see Eq.~\ref{ghxx}).
The diphoton signal rate can be sizably deviate from the SM prediction, either enhanced or
suppressed significantly.
In a large part of the parameter space, the Higgs decay  $h_1\to \chi_{1}^0 \chi_{1}^0$
is accessible at the LHC.

\item  In a major part of the parameter space in the nMSSM,
the decay $h_1\to \chi_{1}^0 \chi_{1}^0$ is dominant over $h_1\to b\bar{b}$
and thus observable at the LHC.
The reason is the lightest neutralino is singlino-like
and is always light, as can be seen from the neutralino mass matrix in Eq.~(\ref{Mx-nMSSM}).
Also, from Eq.~(\ref{ghxx}) we see that the coupling $g_{h_{1} \chi_{1} ^{\rm 0} \chi_{1} ^{\rm 0}}$
can be large (near unity). Due to the new sizable decay $h_1\to \chi_{1}^0 \chi_{1}^0$, the total
width of the SM-like Higgs is greatly enlarged and thus its di-photon signal at the LHC
is severely suppressed. So, if the recently observed di-photon signals at the LHC is verified
in the near future, this model will be excluded.
\end{itemize}

\section{Conclusion}
We examined the Higgs decay to neutralino dark matter in low energy SUSY by
considering three different models: the MSSM, NMSSM and nMSSM.
We considered current experimental constraints
at $2\sigma$ level (including the muon $g-2$ and the dark matter relic density) and
scanned over the parameter space of each model. Then in the allowed parameter space
we calculated
the branching ratio of the SM-like Higgs decay to neutralino dark matter and examined
its observability at the LHC by considering three production channels: the weak boson
fusion $VV\to h$, the associated production with a Z-boson $pp\to hZ+X$ or a pair of top
quarks $pp\to ht\bar{t}+X$.
Our findings are: (i) In the MSSM such a decay  is far below the detectable level;
(ii) In the NMSSM it is accessible in a sizable part of parameter space;
(iii) in the nMSSM it is detectable in a major part of the  parameter space.
(iv) When this invisible decay is sizable, the visible di-photon decay is
suppressed.
So, we conclude that at the LHC the interplay of detecting such an invisible decay
and the visible di-photon decay
may allow for a discrimination of different SUSY models.

\section*{Acknowledgement}
Jingya Zhu thanks Lei Wu for helpful discussion.
This work was supported in part by the National Natural Science Foundation of China
(NNSFC) under grant Nos. 10821504, 11135003, 10775039, 11075045, by Specialized
Research Fund for the Doctoral Program of Higher Education with grant No. 20104104110001, and by the Project of
Knowledge Innovation Program (PKIP) of Chinese Academy of Sciences under grant No.
KJCX2.YW.W10.

\end{document}